\documentclass[nohyper,11pt,letterpaper]{JHEP3}

\pdfoutput=1
\usepackage{amssymb,amsmath}
\usepackage{epsfig}
\usepackage{color}
\usepackage{datetime}

\newcommand\fverb{\setbox\fverbbox=\hbox\bgroup\verb}
\newcommand\fverbdo{\egroup\medskip\noindent%
			\fbox{\unhbox\fverbbox}\ }
\newcommand\fverbit{\egroup\item[\fbox{\unhbox\fverbbox}]}
\newbox\fverbbox

\newcommand{\be}{\begin{equation}}
\newcommand{\ee}{\end{equation}}
\newcommand{\ba}{\begin{eqnarray}}
\newcommand{\ea}{\end{eqnarray}}

\newcommand{\pa}{\partial}

\newcommand{\lap}{\bigtriangleup}

\newcommand{\eq}[1]{(\ref{#1})}

\newcommand{\hh}{\, ,\hspace{1cm}}
\newcommand{\hhh}{\, ,\hspace{0.4cm}}

\newcommand{\ins}[1]{{\mbox{\tiny #1}}}

\newcommand{\inds}[1]{{\scriptscriptstyle #1}}

\title{Bi-conformal symmetry and static Green functions in the
Schwarzschild-Tangherlini spacetimes}

\author{Valeri P. Frolov\thanks{E-mail:
vfrolov@ualberta.ca}\, and
Andrei Zelnikov\thanks{E-mail: zelnikov@ualberta.ca}\\
Theoretical Physics Institute, Department of Physics,
University of Alberta, Edmonton, AB, Canada T6G 2E1}

\abstract{
We study static massless minimally coupled scalar field created by a source 
in a static D-dimensional spacetime. We demonstrate that the corresponding 
equation for this field is invariant under a special transformation of the 
background metric. This transformation consists of the static conformal 
transformation of the spatial part of the metric accompanied by a properly 
chosen transformation of the red-shift factor. Both transformations are 
determined by one function $\Omega$ of the spatial coordinates. We show 
that in a case of higher dimensional spherically symmetric black holes 
one can find such a bi-conformal transformation that the symmetry of the 
D-dimensional metric is enhanced after its application. Namely, the metric 
becomes a direct sum of the metric on a unit sphere and the metric of 2D 
anti-de Sitter space. The method of the heat kernels is used to find the 
Green function in this new space, that allows one, after dimensional 
reduction, to obtain a static Green function in the original space of 
the static black hole. The general useful representation of static Green 
functions is obtained in the Schwarzschild-Tangherlini spacetimes of arbitrary 
dimension. The exact explicit expressions for the static Green functions 
are obtained in such metrics for $D\le 6$. It is shown that in the four 
dimensional case the corresponding Green function coincides with the 
Copson solution.

\bigskip

}

\preprint{Alberta Thy 19-14}

\begin{document}

\section{Introduction}

Problem of finding a field, self-energy, and self-force of charged particles
near black holes have been attracting attention for more than 40 years
\cite{Unruh:1976fc,Smith:1980tv,Zelnikov:1983,Zelnikov:1982in},
though the exact solution for the field of a point electric charge in the
Schwarzschild geometry was first obtained much earlier \cite{Copson01031928}.
The recent interest to the problem of a self-force is stimulated by with a study of
the back-reaction of the field on the particle moving near black holes in
connection with the gravitational wave emission by such particles. At the same
time, even if a charged particle is at rest, study of a self-force might be
quite
interesting in connection with different gedanken experiments with black holes.
More recently, several publication discussed higher-dimensional aspects of this
problem. This study was stimulated by general interest to spacetimes and brane
models with large extra dimensions.

In this paper we study a following problem. Suppose a static charged
particle,
supported by some external force, is at rest close to a static black hole.
We would like to know a field created by this charge. For simplicity we assume
that a particle has a scalar charge which is the origin of a (minimally
coupled) massless
scalar field. Such a problem reduces to finding solutions of a special type
elliptic equation satisfying the regularity condition at the horizon and
decreasing at infinity. A standard method is to expand a solution into modes
obtained by proper separation of variables. However, using the corresponding
series representation of the solution, for example, for the calculation of the
self force, requires a lot of work  which includes mode-by-mode subtraction of
the divergences and summation of the obtained series (see e.g.
\cite{Beach:2014aba}). In some special cases it is possible to use
analytical approximations. For example, when a particle is close to the horizon
one can find its field in a local domain in its vicinity by using the Rindler
approximation (see, e.g., \cite{Frolov:2014gla}). However in such an
approximation the size of the domain must be much smaller than the gravitational
radius and this approximation does not allows one to get information about a
global structure of the field and its dependence on the topology of the
horizon.

Fortunately, there are known interesting examples, when the problem can be
solved exactly and analytically. The best known example is a case of an electric
charge near four-dimensional Schwarzschild black hole. This exact solution was 
obtained by Copson
\cite{Copson01031928} and corrected by Linet \cite{Linet:1976sq} who
demonstrated that additional properly chosen zero-mode solution must be added to
the Copson expression in order to guarantee that the black hole is kept
uncharged. The generalization of this solution to the case of a scalar charge
was found in \cite{Linet:1977vv}. Further generalizations to the
Reissner-Nordstr\"{o}m and Kerr black holes can be found in
\cite{Zelnikov:1982in,Zelnikov:1983,Ottewill:2012aj}. Exact higher-dimensional 
solutions for a
field of a static point charge  near extremely charged black holes (or a set of
such black holes) described by Majumdar-Papapetrou metric was found in
\cite{Frolov:2012jj}.

In this paper we propose a  method which allows one to obtain
solutions for
the field of a charge in a variety of new interesting higher-dimensional black
hole metrics. We shall describe this method in detail in the next section,
but it might be useful to explain it without entering into details just now. We
restrict here ourselves by a simplest case of a static point charge of massless
scalar field $\varPhi$ in a D-dimensional static spacetime. Let us write the
metric in the form ($a,b=1,\ldots,D-1$)
\be\label{met}
ds^2=-\alpha^2 dt^2+g_{ab}dx^a dx^b\hhh \partial_t\alpha=0\hhh\partial_t 
g_{ab}=0
\, .
\ee
We consider a static scalar field $\varPhi$ created by a static source $J$ in 
such a space.
Let us write schematically the corresponding equation in the form
\be\label{eq}
\hat{F}\,\varPhi=-4\pi J\, ,
\ee
where $\hat{F}=\hat{F}_{[g,\alpha]}$ is a corresponding
second order operator
with coefficients depending on the
$(D-1)$-dimensional metric $g_{ab}$ and the red-shift factor $\alpha$. We shall 
show that
the form of this equation remains the same under rescaling
\be\label{scal}
g_{ab}=\Omega^2 \bar{g}_{ab}\hh \alpha=\Omega^{k_\alpha}\bar{\alpha}\hh J=
\Omega^{k_J}\bar{J}\, ,
\ee
where $\Omega$ is a function of spatial coordinates $x^a$ and $k_\alpha$ and 
$k_J$ are properly chosen constants.

Such a transformation is a special case of so-called bi-conformal 
transformations\cite{GarciaParrado:2003hu,GómezLobo20061069}. In our case this 
is a usual conformal transformation of the
$(D-1)$-dimensional metric $g_{ab}$ accompanied by a properly chosen rescaling
of the red-shift factor $\alpha$. One can naturally interpret $\alpha$ as a
$(D-1)$-dimensional dilaton field.
Let us notice that the transformation \eq{scal} `maps' the $D$-dimensional
metric to another one, which in a general case is not a solution of the 
Einstein 
equations. However in special cases within this family of metrics, parametrized 
by a function $\Omega$, there may exist a metric which possesses an enhanced 
symmetry. This symmetry may make it possible to obtain the exact Green function 
in this new $D$-dimensional spacetime and to solve the problem \eq{eq}. In this 
paper we demonstrate how this method works for a static Green 
function in
the Schwarzschild-Tangherlini black-hole spacetimes and discuss its possible
generalizations.

%%%%%%%%%%%%%%%%%%%%%%%%%%%%%%%%%%%%%%%%%%%%%%%%%%%%

\section{Bi-conformal symmetry}

\subsection{Static Green function}

The action for the minimal scalar field $\varPhi(X)$ is
\be
I=-{1\over 8\pi}\int
dX\,\sqrt{-g^{\ins{D}}}\,\varPhi^{;\mu}\varPhi_{;\mu}+\int
dX\,\sqrt{-g^{\ins{D}}}\, J \varPhi\,.
\ee
Here $J(X)$ is a scalar charge density and $\sqrt{-g^{\ins{D}}}$ is the square 
root of the determinant of the $D$-dimensional metric $g^{\ins{D}}_{\mu\nu}$.
The scalar field obeys the equation
\be\label{BoxPhi}
\Box\,\varPhi=-4\pi J\, ,
\ee
where $\Box$ is the standard covariant $D$-dimensional D'Alembertian.
The solution of this equation can be written in terms of the retarded
Green function $\mathbb{G}_\ins{Ret}(X,X')$
\be
\varPhi(X)=4\pi\int
dX'\sqrt{-g^{\ins{D}}(X')}\,\mathbb{G}_\ins{Ret}(X,X')J(X')\,.
\ee
The retarded Green function satisfies the equation
\be
\Box 
\mathbb{G}_\ins{Ret}(X,X')=-\delta(X,X')=-{\delta(X-X')\over\sqrt{-g^{\ins{D}}}}
\,.
\ee

Let us assume that the spacetime is static with the metric (\ref{met}), so 
that $X=(t,x^a)$ and
\be
\delta(X,X')=-{ \delta(t-t')\delta(x-x')\over
\alpha(x')\sqrt{g(x')}}\,.
\ee
The equation (\ref{BoxPhi}) for a static field $\varPhi$ created by a static 
source $J$ takes the form (\ref{eq}) with
\be
\hat{F}={1\over\alpha\sqrt{g}}\partial_a \left(\alpha\sqrt{g} 
g^{ab}\partial_b\right)\, .
\ee
A solution of this equation can be written in the form
\be
\varPhi(x)=4\pi\int dx'\alpha(x')\sqrt{g(x')}\,{G}(x,x')J(x'),
\ee
where ${G}(x,x')$ is a static Green function. The latter is a solution of
the equation
\be\label{FG}
\hat{F}\,{G}(x,x') = -{\delta(x-x')\over\alpha\sqrt{g}}\, .
\ee
The current of a static point charge $q$ located at the point $y$ reads
\be
J(x)=q{\delta(x-y)\over \sqrt{g}}\,.
\ee
The scalar field at the point $x$ created by this charge is 
\be\label{potential}
\varPhi(x)=4\pi q\,\alpha(y)\,{G}(x,y)\,.
\ee
The $D$-dimensional Green function $\mathbb{G}_\ins{Ret}(X,X')$ in a static 
spacetime depends on $t-t'$ and a static Green function can be written as the 
time 
integral
\be\label{Gret}
{G}(x,x')=\int_{-\infty}^{\infty} dt\,\mathbb{G}_\ins{Ret}(t,x;0,x')\, .
\ee

\subsection{Bi-conformal transformation of higher-dimensional static 
spherically 
symmetric spacetimes and symmetry enhancement}

Now let us consider a bi-conformal transformation (\ref{scal})
of the static metric \eq{met}. It is easy to check that the equation 
(\ref{eq}) preserves its form under this transformation provided \be
k_\alpha=-(D-3)\hh k_J=2\, .
\ee
This bi-conformal invariance of the static field equation was used earlier for
calculations of a self-energy and a self-force of charges in
various gravitational backgrounds
\cite{Frolov:2012jj,Frolov:2012zd,Frolov:2012xf,Frolov:2012ip,Frolov:2013qia,
Frolov:2014gla}. This symmetry enables one to relate static Green functions for
the fields in different spacetimes, for example, in the geometry of a set 
of
extremely charged black holes (the Majumdar-Papapetrou spacetime) and in the
flat Minkowski spacetime.

We focus now on a special class of metrics, namely static spherically symmetric 
metrics of the form ($n=D-3$)
\be\label{ds}
ds^2=-f(r)\,dt^2+f^{-1}(r)\,dr^2+r^2\,d\omega^2_{n+1}\hh
f=1-{2M\over r^{n}}+{Q^2\over r^{2n}}\, ,
\ee
where $d\omega_{n+1}^2$ is the line
element on the $(n+1)$-dimensional unit sphere.
This is a metric of a charged $D$-dimensional black hole.

Let us make the bi-conformal transformation of this metric by using the 
conformal function
\be\label{Omega00}
\Omega=r/a\, ,
\ee
where $a$ is an arbitrary constant of a dimension of length. 
The transformed metric takes the form of a direct sum of two metrics
\be
d\bar{s}^2=dh^2+a^2 \,d\omega^2_{n+1}\hhh
dh^2=-\left({r\over a}\right)^{2n}f(r)\,dt^2+{a^2\over r^2 
f(r)}\,dr^2\, .
\ee
It is easy to check that Ricci scalar $\bar{R}$ for the 2D metric $dh^2$ is 
constant
\be
\bar{R}=-2n^2/a^2\, .
\ee
This means that this 2D space is homogeneous. In fact this is a 2D 
anti-de Sitter 
metric. The symmetry $R^1\times SO(n+2)$ of original spacetime after the 
bi-conformal transformation is enhanced and becomes $SO(1,2)\times 
SO(n+2)$.

In this paper we shall demonstrate how to use this enhanced symmetry in order 
to 
obtain a static Green in the Schwarzschild-Tangherlini metric describing a 
neutral static higher-dimensional black hole. The method does work for a case 
of higher-dimensional charged black holes as well, but the required 
calculations are more complicated and we discuss this case in another paper.

%%%%%%%%%%%%%%%%%%%%%%%%%%%%%%%%%%%%%%%%%%%%%%%%%%%%%%%%%%%%%

\section{Scalar charges near static vacuum black hole}

\subsection{Schwarzschild-Tangherlini metric}

The Schwarzschild-Tangherlini  metric is a special case of \eq{ds} with $Q=0$. It 
can be written as follows
\be\begin{split}\label{S-T}
ds^2&=-f\,dt^2+f^{-1}\,dr^2+ r^2\, d\omega_{n+1}^2\,,   \\
f&=1-\left({r_\ins{g}\over r}\right)^n
\hh r_\ins{g}=(2M)^{1/n}
\hh n=D-3\,.
\end{split}\ee
Here $r_\ins{g}$ is the gravitational radius. The line
element on the $(n+1)$-dimensional unit sphere $d\omega_{n+1}^2$ is defined by the 
following recursive relations
\be
d\omega_{n+1}^2=d\theta_{n}^2+\sin^2\theta_{n}\,d\omega_{n}^2\hh
d\omega_{0}^2 = d\phi^2\, .
\ee
We denoted $\theta_0\equiv\phi\in[0,2\pi]$. All other coordinates 
$\theta_{i>0}\in[0,\pi]$.

The static Green function in the 
Schwarzschild-Tangherlini spacetime \eq{S-T} satisfies \eq{FG} with
$\alpha=\sqrt{f}$. It is the solution of the equation
\be\label{lapGreen}
\lap{G}(x,x')=-{\delta(r-r')\delta^{n+1}(\omega ,\omega')\over r^{n+1}}\hh
\lap={1\over r^{n+1}}\left({\partial_r}\,r^{n+1} f\,{\partial_r}\right)
+{1\over r^{2}}\,\lap^{n+1}_{\omega}\, .
\ee
Here $\lap^{n+1}_{\omega}$ and $\delta(\omega ,\omega')$ are the Laplace 
operator and a covariant delta-function on the unit $(n+1)$-dimensional sphere, 
respectively, 
\be\begin{split}
\lap^{n+1}_{\omega}&=\partial^2_{\theta_{n}}+n{\cos\theta_{n}
\over\sin\theta_{n}}
\partial_{\theta_{n}}+{1\over\sin^2\theta_{n}}\lap^{n}_{\omega}\hh
\lap^{1}_{\omega}=\partial^2_{\phi}\, ,\\
\delta^{n+1}(\omega ,\omega')&={\delta(\theta_{n}-\theta'_{n})\over 
\sin^n\theta_{n}}\delta^{n}(\omega ,\omega')\hh
\delta^{1}(\omega ,\omega')=\delta(\phi-\phi')\, .
\end{split}\ee

It is convenient to introduce a new radial variable $\rho$ related to the
Schwarzschild radial coordinate $r$ as follows
\be
{ r \over r_\ins{g}}=\left({\rho+1\over 2}\right)^{1/n}\hhh \rho=2\,{r^n \over
r^n_\ins{g}}-1\, .
\ee
The Tangherlini metric takes the form
\be\label{Tangherlini}
ds^2=-{\rho-1\over\rho+1}\,dt^2+ r_g^2\left({\rho+1\over
2}\right)^{2/n}\left[ {1\over n^2(\rho^2-1)}\,d\rho^2+d\omega^2_{n+1} \right]
\,.
\ee
Using relations
\be
{\delta(r-r')\over r^{n-1}}={2n\over r^n_\ins{g}}\,\delta(\rho-\rho')
\hh
\partial_r={2n\over r_\ins{g}}\,\left({\rho+1\over
2}\right)^{(n-1)/n}\,\partial_\rho
\,,
\ee
one can show that the equation \eq{lapGreen}  for the static Green function
takes the form
\be\label{TangherliniLap}
\left[n^2(\rho^2-1)\,\partial^2_\rho+2n^2\rho\,\partial_\rho+\lap_{\omega 
}^{n+1}
\right] G(x,x')=-{2n\over r_g^{n}}\delta(\rho-\rho')\delta^{n+1}(\omega 
,\omega')
\,.
\ee
The scalar field $\varPhi(x)$ of a point charge $q$ located at $y$ is given by 
\eq{potential}.

%%%%%%%%%%%%%%%%%%%%%%%%%%%%%%%%%%%%

\subsection{Boundary conditions}

The horizon, where $\rho=1$, is a singular point of the equation %%
\eq{TangherliniLap}. In order to uniquely specify a solution of this 
equation one needs, besides a natural condition of decreasing of a solution at 
infinity, to impose specially chosen boundary condition at $\rho=1$. This 
condition follows from the requirement that the field is regular at the horizon 
in regular coordinates that cover it. Let us discuss this point in more detail.

In a physical set up, our problem can be formulated as follows. Let a 
spherically symmetric black hole be formed by a collapse of a spherical object. 
The metric outside this object is always (before and after black hole 
formation) 
described by the Schwarzschild-Tangherlini  solution. We assume that neither 
before nor after the collapse there are no free waves of the scalar field. The 
form 
of 
the solution \eq{Gret} reflects this. Let us consider now a static charge in 
the spacetime of the eternal black hole with the same parameters. The late time 
solution for the field $\varPhi$ is the same in both cases (eternal black hole 
and a black hole created by a collapse). Thus, one can start with a 
solution $\varPhi$ for a static source $J$ placed near the eternal black hole. 
By causality, this solution is defined only out of the past horizon $H^-$. 
Beyond this surface it can have an arbitrary continuation. However, in order to 
escape singularity on $H^-$ one needs to assume a special continuity property 
of $\varPhi$. This problem and the how to deal with it was 
discussed in \cite{Zelnikov:1982in,DemiansiNovikov1982,FrolovNovikov1998}. 
Because the source $J$ is static, there is no radiation and, hence, radiation 
reaction terms and the field $\varPhi$ in the black hole exterior 
remain the same. The 
natural way is to consider a symmetric in time reflection part 
\cite{Poisson:2011nh} of 
$\mathbb{G}_\ins{Ret}$. At the same time, this solution in the global spacetime 
of the eternal black hole becomes symmetric with respect to the time 
reflection. In particular, the value of the field on the horizons $H^{\pm}$, 
that are  
functions of the angle variables, are the same and coincide with the value of 
the field on the bifurcation surface of the horizons. As the result, 
one can analytically continue the static field $\varPhi$ to the whole 
spacetime of the eternal black hole. After the Wick rotation, the static field 
$\varPhi$ is to be regular at the 
Euclidean horizon. This property can be used to obtain the proper boundary 
conditions for a static equation \eq{TangherliniLap} at its singular point 
$\rho=1$.

Let us make the Wick rotation $t=it_\ins{E}$ in the metric \eq{Tangherlini} 
and consider the vicinity of the Euclidean horizon. Let us write $\rho$ in the 
form 
\be
\rho=1+{n^2\over 2r_g^2}\ell^2+\ldots\, . 
\ee
$\ell=0$ is a position of the horizon and $\ell$ is a proper length distance 
from it. Here $(\ldots)$ denotes terms of higher order in $\ell^2$. The 
Euclidean 
version of the metric \eq{Tangherlini} takes the form 
\be
ds_E^2\approx {n^2\over 4r_g^2} \ell^2 dt_\ins{E}^2+d\ell^2 +r_g^2 
d\omega_{n+1}^2\, .
\ee
The regularity of the Euclidean horizon requires that the coordinate 
$t_\ins{E}$ 
is 
periodic with the period 
\be\label{beta}
\beta={2\pi\over \kappa}={4\pi r_\ins{g}\over n}\,,
\ee
where $\kappa$ is the surface gravity 
\be\label{surface_grav}
\kappa={n\over 2r_\ins{g}}\, .
\ee

The operator $\hat{F}$ near the horizon is
\be
\hat{F}\approx \left( \partial_{\ell}^2+{1\over 
\ell}\partial 
_{\ell}\right) +{1\over r_\ins{g}^2}\lap_{\omega}^{n+1}\, .
\ee
Let us write a solution of the equation
\be\label{eqq}
\hat{F}\varPhi=0
\ee
near the horizon in the form
\be
\varPhi=f_0+f_1\ell+f_2\ell^2+\ldots\, ,
\ee
where $f_i$ are functions on a unit sphere.
Then \eq{eqq} implies
\be
f_1=0\hh f_2=-{1\over 4r_\ins{g}^2}\lap_{\omega}^{n+1}f_0\, .
\ee
This means that $\varPhi$ is a regular function near the Euclidean horizon. This is 
a proper boundary condition for $\varPhi$ near $\rho=1$.

For any solution of \eq{TangherliniLap} one can always add a solution of a 
homogeneous equation with the same operator. This arbitrariness is related to 
the zero-modes contribution to the Green function. These extra zero-mode terms 
in 
the Green function have to be symmetric in $x$ and $x'$ and regular outside the 
horizon. In our case of the \eq{TangherliniLap} they are characterized by two 
arbitrary constants $C_1$ and $C_2$
\be
C_1\ln\left({\rho-1\over\rho+1}\right)\ln\left({\rho'-1\over\rho'+1}
\right)+C_2\left[
\ln\left({\rho-1\over\rho+1}\right)+\ln\left({\rho'-1\over\rho'+1}\right)\right]
\,.
\ee
Using these terms one can  adjust the solution for the scalar field to make it 
regular on the horizon and satisfy proper boundary conditions at infinity.

Let us notice that if one makes a bi-conformal transformation with $\Omega$ 
regular on the horizon $\Omega=\Omega_0+\Omega_2\ell^2+\ldots$, then the 
surface gravity $\kappa$ changes as follows
$\bar{\kappa}=\Omega_0^{n+1}\kappa$.

%%%%%%%%%%%%%%%%%%%%%%%%%%%%%%%%%%%%%%%%%%%%%%%%%%%%%%%%%%%%%
%%%%%%%%%%%%%%%%%%%%%%%%%%%%%%%%%%%%%%%%%%%%%%%%%%%%%%%%%%%%%

\section{Scalar charges in the Bertotti-Robinson
spacetime}

Let us make a bi-conformal transformation of the Tangherlini metric 
\be
g_{ab}=\Omega^2\bar{g}_{ab}\hh \alpha=\Omega^{-n}\bar{\alpha}\,,
\ee
by choosing
the conformal factor \eq{Omega00} with $a=r_\ins{g}$
\be\label{Omega}
\Omega={r\over r_\ins{g}}=\left({\rho+1\over 2}\right)^{1/n}.
\ee
Notice that at the horizon $\Omega=1$ so that this transformation does not 
modify the surface gravity \eq{surface_grav}. The Tangherlini metric
\eq{Tangherlini} transforms to 
\be\label{Bertotti-Robinson}
d\bar{s}^2=\left({r_\ins{g}\over n}\right)^2
\left[-\kappa^2(\rho^2-1)\,dt^2+(\rho^2-1)^{-1}\,d\rho^2\right] 
+r_\ins{g}^2\,d\omega_{n+1}^2
\,.
\ee
Its Euclidean version is the metric on $H^2\times S^{n+1}$ and has the form
\be\label{Bertotti-Robinson_E}
d\bar{s}_\ins{E}^2=\left({r_\ins{g}\over n}\right)^2
\left[\kappa^2(\rho^2-1)\,dt_\ins{E}^2+(\rho^2-1)^{-1}\,d\rho^2\right] 
+r_\ins{g}^2\,d\omega_{ n+1 }^2
\,.
\ee
Let us notice that the metric \eq{Bertotti-Robinson} is a special case of the 
Bertotti-Robinson metric
\be
ds^2=b^2\,dH^2+a^2\, d\omega_{ n+1 }^2
\ee
with $b=r_\ins{g}/n$ and $a=r_\ins{g}$. Here
\be
dH^2=-(\rho^2-1)\,d\bar{\sigma}^2+(\rho^2-1)^{-1}\,d\rho^2
\ee
is the metric of the two-dimensional AdS spacetime of a unit curvature radius.
The coordinate $\bar{\sigma}$ is related to the Schwarzschild time as follows
$\bar{\sigma}=\kappa t$.

After the Wick rotation $\bar{\sigma}=i\sigma$ this metric takes the form
\be\label{Bertotti-Robinson_EE}
dH^2_\ins{E}=(\rho^2-1)\,d\sigma^2+(\rho^2-1)^{-1}\,d\rho^2\,.
\ee
It is regular at the Euclidean horizon $\rho=1$ provided $\sigma$ is 
periodic with the period $2\pi$. The scalar curvatures of the sphere $S^{n+1}$ 
of the radius $a$ and of the
hyperboloid $H^2$ of the radius $b$ are
\be
R_\inds{S^{n+1}}={n(n+1)\over a^2}
\hh
R_\inds{H^2}=-{2\over b^2}\,.
\ee

Denote by ${\mathbb G}$ the Euclidean Green function of the operator 
$\bar{\Box}_\ins{E}$ defined on the Euclidean Bertotti-Robinson 
metric \eq{Bertotti-Robinson_E}. It is the solution of the equation
\be
\bar{\Box}_\ins{E}\,{\mathbb G}(X_\ins{E},X'_\ins{E}
)=-\delta(X_\ins{E},X'_\ins{E})\,,
\ee
decreasing at infinity and regular at the Euclidean horizon $\rho=1$. We define 
the
static Green function $G^\ins{BR}(x,x')$ as
\be\begin{split}\label{GBR}
G^\ins{BR}(x,x')&=\int_0^{\beta} dt_\ins{E}\,{\mathbb G}(t_\ins{E},x;0,x')\,.
\end{split}\ee

Then the static Green function of the minimally coupled massless scalar 
field on the Bertotti-Robinson spacetime can be calculated as the 
integral of the Euclidean Green function of the scalar field 
in the Bertotti-Robinson spacetime. It is regular at 
the horizon, decreases at infinity, and satisfies the equation
\be\label{BertottiLap}
\left[n^2(\rho^2-1)\,\partial^2_\rho+2n^2\rho\,\partial_\rho+\lap_{\omega}^{n+1}
\right] G^\ins{BR}(x,x')=-{2n\over 
r_\ins{g}^{n}}\delta(\rho-\rho')\delta^{n+1}(\omega 
,\omega')\,,
\ee
which coincides identically with the equation \eq{TangherliniLap} for the 
static 
Green function on the Schwarzschild-Tangherlini background.
This means that 
\be\label{GGBR}
G^\ins{BR}(x,x')=G(x,x')\,,
\ee
where $G(x,x')$ is the static Green 
function for the original Schwarzschild-Tangherlini spacetime.

%%%%%%%%%%%%%%%%%%%%%%%%%%%%%%%%%%%%%%%%

\section{Green functions and heat kernels}

\subsection{General relations}

We demonstrated that the static Green function in the 
Schwarzschild-Tangherlini spacetime can be found as soon as the corresponding 
Green function for the Bertotti-Robinson geometry is known. For this purpose it 
is convenient to use the heat kernel technique. Namely, the Euclidean Green 
function for the operator ${\Box}_\ins{E}$ is
\be
{\mathbb 
G}(X_\ins{E},X'_\ins{E})=\int_0^{\infty}ds\,K(s|X_\ins{E},X'_\ins{E})\,.
\ee
Here the heat kernel $K(s|X_\ins{E},X'_\ins{E})$ is the solution of the problem
\be
(\pa_s-{\Box}_\ins{E})\,K(s|X_\ins{E},X'_\ins{E})=0
\hh 
K(0|X_\ins{E},X'_\ins{E})=\delta(X_\ins{E},X'_\ins{E})
\,,
\ee
which satisfies the same boundary conditions with respect to its arguments 
$X_\ins{E}$ 
and $X'_\ins{E}$ as the Green function in question.

The Bertotti-Robinson metric \eq{Bertotti-Robinson_EE} is the direct sum of two 
geometries $H^2$ and $S^{n+1}$. This implies that the  heat kernel $K$ is the 
product of the heat kernels $K_\inds{H^2}$ and $K_\inds{S^{n+1}}$ on the 
hyperboloid $H^2$ and on the sphere $S^{n+1}$, correspondingly. Both spaces are 
homogeneous and isotropic and the corresponding heat kernels are known 
explicitly \cite{Camporesi:1990wm}.

%%%%%%%%%%%%%%%%%%%%%%%%%%%%%

\subsection{Heat kernel on $H^2$}

The heat kernel on $H^2$ of the radius $b$ 
\be\label{H2}
dH_\ins{E}^2=b^2\left[(\rho^2-1)\,d\sigma^2+(\rho^2-1)^{-1}\,d\rho^2\right]
\ee
reads \cite{Camporesi:1990wm}
\be
K_\inds{H^2}(s|\chi)={\sqrt{2} b \over (4\pi
s)^{3/2}}\,e^{-s/(4b^2)}\int_{\chi}^{\infty}dy\,{y\,e^{-b^2
y^2/(4s)}\over(\cosh y-\cosh(\chi))^{1/2}}
\,.
\ee
Here $\chi$ is the geodesic distance between
two points on the $H^2$ of the unit radius $b=1$. It can be found from the 
relation
\be\begin{split}\label{chi}
\cosh(\chi)&=\rho\rho'-\sqrt{\rho^2-1}\sqrt{\rho'{}^2-1}\cos(\sigma-\sigma')\,.
\end{split}\ee

%%%%%%%%%%%%%%%%%%%%%%%%%%%%%

\subsection{Heat kernel on $S^{n+1}$}

The heat kernel on $S^2$ of the radius $a$ reads \cite{Camporesi:1990wm}
\be
K_\inds{S^2}(s|\gamma)={\sqrt{2} a \over (4\pi
s)^{3/2}}\,e^{s/(4 a^2)}\sum_{k=-\infty}^{\infty}(-1)^k
\int_{\gamma}^{\pi} d\phi {(\phi+2\pi
k)\,e^{-a^2(\phi+2\pi k)^2/(4s)}\over(\cos\gamma-\cos\phi)^{1/2}}
\,.
\ee
Another equivalent representation of this kernel is
\be
K_\inds{S^2}(s|\gamma)={1\over 4\pi a^2}
\sum_{l=0}^{\infty}(2l+1)P_l(\cos\gamma)\,e^{-{sl(l+1)\over a^2}} \,.
\ee
Here $\gamma$ is the
geodesic distance between two points on the unit $S^2$ ($a=1$)
\be
\cos\gamma=\cos(\theta_1)\cos(\theta'_1)
+\sin(\theta_1)\sin(\theta'_1)\cos(\phi-\phi')\,.
\ee

The heat kernel on $S^3$ of the radius $a$ reads \cite{Camporesi:1990wm}
\be
K_\inds{S^3}(s|\gamma)={1\over (4\pi
s)^{3/2}}\,e^{s/a^2}\sum_{k=-\infty}^{\infty}{(\gamma+2\pi k)
\,e^{-a^2(\gamma+2\pi k)^2/(4s)}\over\sin\gamma}\,,
\ee
where $\gamma$ is the geodesic distance between two points on the unit $S^3$ 
($a=1$)
\be
\cos\gamma=\cos(\theta_2)\cos(\theta'_2)+\sin(\theta_2)\sin(\theta'_2)[
\cos(\theta_1)\cos(\theta'_1)+\sin(\theta_1)\sin(\theta'_1)\cos(\phi-\phi')]\,.
\ee

The heat kernels on $S^{n+1}$ can be derived from $K_\inds{S^2}$ and
$K_\inds{S^3}$using the relations  (see \cite{Camporesi:1990wm}
Eq.(8.12)-Eq(8.13))
\be
K_\inds{S^{n+1}}(s|\gamma)=e^{(n^2-1)s\over 4 a^2}\left({1\over
2\pi a^2}{\partial\over \partial \cos\gamma}\right)^{(n-1)\over
2}K_\inds{S^2}(s|\gamma)\hh n~~\mbox{odd} \,,
\ee
\be
K_\inds{S^{n+1}}(s|\gamma)=e^{(n^2-4) s\over 4 a^2}\left({1\over
2\pi a^2}{\partial\over \partial \cos\gamma}\right)^{(n-2)\over
2}K_\inds{S^3}(s|\gamma)\hh n~~\mbox{even} \,.
\ee

%%%%%%%%%%%%%%%%%%%%%%%%%%%%%

\section{Results}

Combining the results of the previous section together one obtains
\be
K(s|X_\ins{E},X'_\ins{E})={K}_{}(s|\chi,\gamma)={K}_{H^2}(s|\chi)\times{K}_{S^{
n+1}}(s|\gamma)\,.
\ee
Because of the symmetry of the Euclidean Bertotti-Robinson space the heat 
kernel 
$K$ can be written as the function of only three variables:  
the proper time $s$ and geodesic distances $\chi$ and $\gamma$ on the unit 
hyperboloid and the unit sphere.

In what follows we use the relation $b=a/n$ and identify $a$ with $r_\ins{g}$. 
For briefness we put $a=r_\ins{g}$ only at the end of calculations.
One should keep in mind that the cases of
spacetimes of even and odd dimensions differ qualitatively, and should be
studied separately.

\subsection{Odd $D$}

For odd $D$ (even $n$)
\be\begin{split}
{K}_{}(s|\chi,\gamma)&={a\over n}\left({1\over 2\pi
a^2}{\partial\over
\partial \cos\gamma}\right)^{(n-2)/2}
{\sqrt{2}\over (4\pi s)^{3}}{1\over\sin\gamma}\\
&\sum_{k=-\infty}^{\infty}{(\gamma+2\pi
k)\,e^{-a^2(\gamma+2\pi k)^2/(4s)}} \int_{\chi}^{\infty}dy\,{y\,e^{-a^2
y^2/(4 n^2 s)}\over(\cosh y-\cosh\chi)^{1/2}} \,.
\end{split}\ee
Equivalently one can write
\be\begin{split}
{K}_{}(s|\chi,\gamma)&={a\over n}\left({1\over 2\pi
a^2}{\partial\over
\partial \cos\gamma}\right)^{n/2}
{\sqrt{2}\over (4\pi s)^{2}}\\
&\sum_{k=-\infty}^{\infty}{e^{-a^2(\gamma+2\pi k)^2/(4s)}}
\int_{\chi}^{\infty}dy\,{y\,e^{-a^2
y^2/(4 n^2 s)}\over(\cosh y-\cosh\chi)^{1/2}} \,.
\end{split}\ee

The Green function is an integral of the heat kernel over $s$
\be
{\mathbb G}_{}(\chi,\gamma)=\int_0^{\infty}ds\,{K}(s|\chi,\gamma) \,.
\ee
For separated points we can perform this integration first
\be
{\mathbb G}_{}(\chi,\gamma)={1\over an}{\sqrt{2}\over 4\pi^2}
\left({1\over 2\pi a^2}{\partial\over \partial \cos\gamma}\right)^{n/2}
\sum_{k=-\infty}^{\infty}
\int_{\chi}^{\infty}dy\,{y\over(\cosh y-\cosh\chi)^{1/2}} {1\over
\left[{y^2\over n^2}+(\gamma+2\pi k)^2\right]} \,.
\ee
Summation over $k$ gives
\be
\sum_{k=-\infty}^{\infty}{1\over \left[{y^2\over n^2}+(\gamma+2\pi k)^2\right]}
={n\over 2 y}\,{\sinh\left({y\over 2n}\right)\cosh\left({y\over 2n}\right)\over
\cosh\left({y\over
2n}\right)^2-\cos\left({\gamma\over 2}\right)^2}
={n\over 2 y}\,{\sinh\left({y\over n}\right)\over \cosh\left({y\over
n}\right)-\cos\gamma} \,.
\ee
Therefore, the $D$-dimensional Euclidean Green function on the 
Bertotti-Robinson spacetime becomes
\be\label{oddG}
{\mathbb G}_{}(\chi,\gamma)={\sqrt{2}\over 8\pi^2 a}
\left({1\over 2\pi a^2}{\partial\over \partial \cos\gamma}\right)^{n/2}
\int_{\chi}^{\infty}dy\,{1\over(\cosh y-\cosh\chi)^{1/2}}\,{\sinh(y/n)\over
\cosh(y/n)-\cos\gamma} \,,
\ee
where $\chi$ is given by \eq{chi}.

This is the Euclidean Green function of a massless
minimally coupled scalar field on the background of the Euclidean
$D$-dimensional (odd $D$) Bertotti-Robinson spacetime. This Green function can
be used, e.g., for computation of the vacuum mean value of the stress-energy
tensor or $\langle\varPhi^2\rangle$ on the higher-dimensional Bertotti-Robinson
geometry. In four dimensions the Green function in closed form was found
in \cite{Kofman:1983nj} and was used for calculation of the vacuum mean value
of the stress-energy tensor in \cite{Ottewill:2012mq}.

Using the properties \eq{GBR},\eq{GGBR} we finally get the static Green function 
on 
the 
odd-dimensional Schwarzschild-Tangherlini spacetimes 
\be\label{GSchw}
{G}(x,x')={2\sqrt{2}\over 8\pi^2 n}
\left({1\over 2\pi a^2}{\partial\over \partial 
\cos\gamma}\right)^{n/2}\,\int_0^{2\pi} d\sigma\,
\int_{\chi}^{\infty}dy\,{1\over(\cosh y-\cosh\chi)^{1/2}}\,{\sinh(y/n)\over
\cosh(y/n)-\cos\gamma} \,,
\ee
where $x=(\rho,\theta_i)$, $a=r_\ins{g}$, and
\be\begin{split}
\cosh(\chi)&=\rho\rho'-\sqrt{\rho^2-1}\sqrt{\rho'{}^2-1}\cos\sigma
\hh \rho=2\,{r^n \over r^n_\ins{g}}-1
 \,.
\end{split}\ee

%%%%%%%%%%%%%%%%%%%%%%%%%%%%%%

\subsection{Example. $D=5$}\label{5D}

In five dimensions ($n=2$) we have
\be\begin{split}
{\mathbb G}(\chi,\gamma)&={\sqrt{2}\over 8\pi^2 a}
\left({1\over 2\pi a^2}{\partial\over \partial \cos\gamma}\right)
\int_{\chi}^{\infty}dy\,{1\over\sqrt{\cosh
y-\cosh\chi}}\,{\sinh(y/2)\over
\cosh(y/2)-\cos\gamma}
\\
&={1\over (2\pi a)^3}
\left({\partial\over \partial \cos\gamma}\right)
\left\{\lambda
{\left[
\arctan\left({\lambda\cos\gamma}\right)
+{\pi\over 2}\right]}
\right\} \,,
\end{split}\ee
where
\be
\lambda=\left(\cosh\left({\chi\over 2}\right)^2-\cos(\gamma)^2\right)^{-1/2}\,.
\ee
Performing the differentiation, we obtain a closed form of the Euclidean Green 
function in the
Euclidean Bertotti-Robinson spacetime \eq{Bertotti-Robinson_E}
\be\begin{split}
{\mathbb G}(\chi,\gamma)&={1\over (2\pi a)^3}
\left\{ {\lambda^3\cos\gamma}
\left[ \arctan\left({\lambda\cos\gamma}\right)
+{\pi\over 2} \right]-\lambda^2 \right\} \,.
\end{split}\ee

The static Tangherlini Green function can be written as follows
\be\begin{split}\label{5DGreen0}
G(x,x')&=a\,\int_0^{2\pi} d\sigma\,{\mathbb G}(\chi,\gamma) \hh 
a=r_\ins{g} \,.
\end{split}\ee
The result of the integration over the Euclidean time $\sigma$ can be
expressed in terms of the elliptic functions. Using the properties of the
elliptic functions presented in the Appendix \ref{elliptic} we obtain the
static Green function in the five-dimensional Bertotti-Robinson spacetime
\be\begin{split}\label{5DGreen}
G(x,x')&={1\over 4\pi^2 r_\ins{g}^2}{1\over(\rho^2-1)^{1/4}(\rho'{}^2-1)^{1/4}}
\left({\partial\over \partial \cos\gamma}\right)
\,\varkappa\,\left[\mathbf{F}\left(\psi,\varkappa\right)+\mathbf{K}
\left(\varkappa\right)\right]\,,
\end{split}\ee
where
\be\begin{split}
\sin\psi&=\cos\gamma\,{\sqrt{2}\over
\sqrt{\rho\rho'-\sqrt{\rho^2-1}\sqrt{\rho'{}^2-1}+1}}  \,,
\\
\varkappa&={\sqrt{2}\,(\rho^2-1)^{1/4}(\rho'{}^2-1)^{1/4}
\over\sqrt{\rho\rho'+\sqrt{\rho^2-1}\sqrt{\rho'{}^2-1}+1-2\cos^2\gamma}} \,.
\end{split}\ee
In order to rewrite the result in terms of the Schwarzschild  radial coordinate 
$r$ one has to substitute here
\be
\rho=2\,{r^2 \over r^2_\ins{g}}-1 \,,
\ee
In the limit when $\rho=\rho'$ one gets
\be
\sin\psi=\cos\gamma \hh 
\varkappa=\sqrt{\rho^2-1\over\rho^2-\cos^2\gamma} \,.
\ee

%%%%%%%%%%%%%%%%%%%%%%%%%%%%%%%%%%%%%%%%

\subsection{Even $D$}

In even dimensions (odd $n$) the heat kernel has the form
\be\begin{split}
{K}_{}(s|\chi,\gamma)&=-4{a^2\over n}\left({1\over 2\pi
a^2}{\partial\over
\partial \cos\gamma}\right)^{(n+1)/2}
{1\over (4\pi s)^{2}}\sum_{k=-\infty}^{\infty}(-1)^k\\
&\times\int_{\chi}^{\infty}dy\,{y\,e^{-a^2 y^2/(4 n^2 s)}\over(\cosh
y-\cosh\chi)^{1/2}}\int_{\gamma}^\pi
d\phi(\cos\gamma-\cos\phi)^{1/2}{\partial\over\partial\phi} e^{-{a^2(\phi+2\pi
k)^2\over 4 s}} \,.
\end{split}\ee
The Green function reads
\be\begin{split}
&{\mathbb G}_{}(\chi,\gamma)=-{1\over n}{1\over \pi^2}
\left({1\over 2\pi a^2}{\partial\over \partial \cos\gamma}\right)^{(n+1)/2}
\sum_{k=-\infty}^{\infty}(-1)^k\\
&\times\int_{\chi}^{\infty}dy\,{y\over(\cosh y-\cosh\chi)^{1/2}}
\int_{\gamma}^\pi
d\phi\,(\cos\gamma-\cos\phi)^{1/2}{\partial\over\partial\phi}{1\over
\left[{y^2\over n^2}+(\phi+2\pi k)^2\right]} \,.
\end{split}\ee
The series can be computed by using the relation
\be
\sum_{k=-\infty}^{\infty}{(-1)^k\over \left[{y^2\over n^2}+(\phi+2\pi
k)^2\right]}
={n\over 2 y}\,{\sinh\left({y\over 2n}\right)\cos\left({\phi\over
2}\right)\over
\cosh\left({y\over
2n}\right)^2-\cos\left({\phi\over 2}\right)^2} \,.
\ee
Then straightforward integration leads to the Euclidean Green function in the 
form
\be
{\mathbb G}_{}(\chi,\gamma)={1\over 4\pi}
\left({1\over 2\pi a^2}{\partial\over \partial \cos\gamma}\right)^{(n+1)/2}
\,A_n \,,
\ee
where
\be
A_n=\int_{\chi}^{\infty}dy\,{1\over\sqrt{\cosh\left({y\over
2}\right)^2-\cosh\left({\chi\over
2}\right)^2}}\,\sinh\left({y\over 2n}\right)
\left[
{\cosh\left({y\over 2n}\right)\over
\sqrt{\cosh\left({y\over 2n}\right)^2-
\cos\left({\gamma\over 2}\right)^2}}-1
\right] \,.
\ee
Because there is always at least one derivative
over $\gamma$ acting on $A_n$, for $n\ge 3$ one can safely omit $-1$ in the
square brackets in the definition of $A_n$. Therefore, for $n\ge 3$ we get
\be\begin{split}
A_n&=\int_{\chi}^{\infty}dy\,{1\over\sqrt{\cosh\left({y}\right)-\cosh\left({
\chi}\right)}}\,
{\sinh\left({y\over n}\right)\over
\sqrt{\cosh\left({y\over n}\right)-
\cos\left({\gamma}\right)}} \,.
\end{split}\ee

Because the static Green functions in the Euclidean Bertotti-Robinson spacetime
and in the Euclidean Tangherlini spacetime coincide, one can use the relations 
\eq{GBR},\eq{GGBR} to get the static Green function on the 
even-dimensional Schwarzschild-Tangherlini spacetimes
\be
{G}(x,x')={a\over 2\pi n}
\left({1\over 2\pi a^2}{\partial\over \partial \cos\gamma}\right)^{(n+1)/2}
\,\int_0^{2\pi}d\sigma\,A_n \,,
\ee
where $x=(\rho,\theta_i)$, $a=r_\ins{g}$, and
\be\begin{split}
\cosh(\chi)&=\rho\rho'-\sqrt{\rho^2-1}\sqrt{\rho'{}^2-1}\cos\sigma
\hh \rho=2\,{r^n \over r^n_\ins{g}}-1 \,.
\end{split}\ee

%%%%%%%%%%%%%%%%%%%%%%%%%%%%%%

\subsection{Example. $D=4$}

In four dimensions ($n=1$) we obtain

\be
A_1=\ln\left({\cosh\left(\chi\right)+1\over
\cosh\left(\chi\right)-\cos\left(\gamma\right)} \right) \,,
\ee
\be
{\mathbb G}(\chi,\gamma)={1\over 4\pi}
\left({1\over 2\pi a^2}{\partial\over \partial \cos\gamma}\right)
\,A_1={1\over 8\pi^2 a^2}\,
{1\over \cosh\left(\chi\right)-\cos\left(\gamma\right)} \,.
\ee

The static Green function is obtained by substitution of \eq{chi},
and integration over $\sigma$ 
\be\begin{split}
G(x,x')&=2a\,\int_0^{2\pi} d\sigma\,{\mathbb G}(\chi,\gamma)\\
&={1\over 4\pi^2 a}\int_0^{2\pi}
d\sigma\,
{1\over
\rho\rho'-\sqrt{\rho^2-1}\sqrt{\rho'{}^2-1}\cos(\sigma)-\cos\left(\gamma\right)}
 \,.
\end{split}\ee
The integral can be taken explicitly and we obtain the closed form for the 
static Green function
\be\begin{split}
G(x,x')&={1\over 2\pi a}
{1\over \sqrt{ \rho^2+\rho'{}^2-2\rho\rho'\cos\gamma
-1+\cos^2\gamma }} \,.
\end{split}\ee

In four dimensions the static Green function for the Schwarzschild black hole
of mass $M$ is obtained then by substitutions
\be
a=r_\ins{g}=2M \hh \rho=2{r\over r_g}-1={r\over M}-1 \,.
\ee
In terms of the Schwarzschild coordinates it reads
\be
G(x,x')={1\over 4\pi}
{1\over \sqrt{ (r-M)^2+(r'-M)^2-2(r-M)(r'-M)\cos\gamma
-M^2\sin^2\gamma }} \,.
\ee
This formula exactly reproduces the closed form of the well known result for
the scalar Green function in four-dimensional Schwarzschild geometry
\cite{Linet:1977vv,FrolovZelnikov:1980,Zelnikov:1982in}.

\subsection{Example. $D=6$}

In six dimensions ($n=3$) we obtain
\be\begin{split}
A_3&=\int_{\chi}^{\infty}dy\,{1\over(\cosh
y-\cosh\chi)^{1/2}}\,{\sinh\left({y\over 3}\right)
\over\sqrt{\cosh\left({y\over 3}\right)-
\cos\left({\gamma}\right)}}\\
&=3\int_{\cosh(\chi/3)}^{\infty}dz\,{1\over\sqrt{4z^3-3z-\cosh\chi}}
\,{1\over\sqrt{z-\cos\gamma}} \,.
\end{split}\ee
This integral can be expressed in terms of the elliptic function $\mathbf{F}$
\be
A_3={6\over 
\sqrt{v(w-u)}}\mathbf{F}\left(\arcsin\sqrt{w-u\over 
w},{w(v-u)\over v(w-u)}\right)\,,
\ee
where
\be
p=\cosh(\chi/3)\hhh u=3p-i\sqrt{3p^2-3}\hhh v=3p+i\sqrt{3p^2-3}\hhh 
w=2(p-\cos\gamma)\,.
\ee
Note that $A_3$ is real in spite of the complexity of the functions $u$ and $v$.
The Green function on the Euclidean Bertotti-Robinson space reads
\be
{\mathbb G}_{}(\chi,\gamma)={1\over 16\pi^3 a^4}
\left({\partial\over \partial \cos\gamma}\right)^{2}
\,A_3 \,.
\ee
Then the static Green function in the six-dimensional Schwarzschild-Tangherlini 
spacetime is given by the integral
\be
G(x,x')={1\over 24 \pi^3 a^3}
\left({\partial\over \partial \cos\gamma}\right)^{2}
\,\int_0^{2\pi}d\sigma\, A_3 \,,
\ee
where $a=r_\ins{g}$.

It is problematic to obtain an answer for the Green functions in a closed form
for $D\ge6$. However a simple integral representation is possible in all
dimensions
what may be good enough for some applications like computing of the
self-force and self-energy of scalar charges in the Schwarzschild-Tangherlini
spacetime.

%%%%%%%%%%%%%%%%%%%%%%%%%%%%%%%%%%%%%%

\section{Discussion}

In this paper we discuss properties of a field created by a static source placed
in the vicinity of a static higher-dimensional black hole. We focused on a
static Green function which allows one to find the field of a point charge. If 
the
spacetime has $D$ dimensions, such a $(D-1)$-dimensional static Green function
can be obtained by the dimensional reduction from a Green function in $D$
dimensions. We demonstrated that there exists a group of transformations of the
original $D$ dimensional metric, depending on a function of $(D-1)$ variables,
which preserves the form of a static $(D-1)$ dimensional field equation. This
transformation consists of a conformal transformation of the spatial part of the
metric accompanied by a properly chosen transformation of  a red-shift factor
(i.e. $g_{tt}$ component of the metric). These transformations  change the $D$
dimensional metric. In a general case, after such a transformation, a solution
of the Einstein equations is transformed to the metric that does not satisfy the
latter with any physically meaningful stress-energy tensor. We call these 
transformations {\em bi-conformal}. We demonstrated that
the metrics describing  higher-dimensional static black holes
(Schwarzschild-Tangherlini metrics and their electrically charged
generalizations) possess a remarkable property: Within a family of metrics
related to such black hole metrics by means of  a bi-conformal transformation
there exist special metrics that have enhanced symmetry. Namely,  the symmetry
$SO(D-1)\times R^1$ of the original spacetime is enhanced up to $SO(D-1)\times
SO(1,2)$. The spacetime with this symmetry is a Bertotti-Robinson metric that is
a direct sum of the metrics on a $(D-2)$-dimensional sphere and 2-dimensional
AdS spacetime.   We used the heat kernel method to obtain a useful 
representation for the
Green function in the higher-dimensional Bertotti-Robinson  spacetime and after
its dimensional reduction we found the static Green function in the original
black hole metric. In four and five dimensions we obtained the closed form for 
the static Green functions in terms of the elliptic and elementary functions.

The described method is quite general. We illustrated its application by
considering a minimally coupled scalar massless field from a static point charge
located in the vicinity of the higher-dimensional  Schwarzschild-Tangherlini
black hole. Generalizations to charged black holes and to the case of an
electric point charge, that are rather straightforward, will be discussed
somewhere else. Let us also mention, that if one applies the static Green for
the calculation of the self-energy and self-force for a point charge, one should
first regularize it and subtract local divergences. This procedure depends on
the choice of the background $D$-dimensional metric. As a result, renormalized
self-energy and self-force in two spacetimes connected by a bi-conformal
transformation are not connected by simple rescaling. This phenomenon,which is
somehow similar to famous conformal anomalies, is called a {\em bi-conformal
anomaly} or {\em self-energy anomaly} (see discussion in
\cite{Frolov:2012xf,Frolov:2012ip,Frolov:2013qia}). We are
going to return to this problem again in future works.

%%%%%%%%%%%%%%%%%%%%%%%%%%%%%%%%%%

%%%%%%%%%%%%%%%%%%%%%%%%%%

\acknowledgments{
This work was partly supported  by  the Natural Sciences and Engineering
Research Council of Canada. The authors are also grateful to the Killam Trust
for its financial support.}

%%%%%%%%%%%%%%%%%%%%%%%%%%

\appendix
\section{Useful properties of the elliptic functions}\label{elliptic}

Calculating the static Green function \eq{5DGreen0} in the section \ref{5D} 
we have used the integrals 
\be
\int_0^{\pi/2} dx\,{1\over
\sqrt{1-k^2\sin^2 x}}\arctan\left({1\over \alpha\sqrt{1-k^2\sin^2 x}}\right)
={\pi\over 2}
\mathbf{F}\left(
\arctan\left({1\over\alpha\sqrt{1-k^2}}\right),k
\right)\,,
\ee
\be
\int_0^{\pi/2} dx\,{1\over
\sqrt{1-k^2\sin^2
x}}=\mathbf{F}\left({\pi\over
2},k\right)=\mathbf{K}\left(k\right)\,.
\ee

The Green function \eq{5DGreen0} takes the form
\be\begin{split}
G(x,x')&={1\over 4\pi^2 a^2}
\left({\partial\over \partial \cos\gamma}\right)
\left\{{1\over\alpha\cos\gamma}\left[\mathbf{F}\left(
\phi,k
\right)+\mathbf{K}\left(k\right)\right]\right\}\,,
\end{split}\ee
where
\be
\phi=\arctan\left({1\over\alpha\sqrt{1-k^2}}\right)\,,
\ee
\be
\alpha^2={1\over
2\cos^2\gamma}\left(\rho\rho'-\sqrt{\rho^2-1}\sqrt{\rho'{}^2-1}
+1-2\cos^2\gamma\right)\,,
\ee
\be
k^2=-{2\sqrt{\rho^2-1}\sqrt{\rho'{}^2-1}\over
\rho\rho'-\sqrt{\rho^2-1}\sqrt{\rho'{}^2-1}
+1-2\cos^2\gamma}\,.
\ee
One gets
\be\begin{split}
\sin\phi&={1\over\sqrt{1+\alpha^2(1-k^2)}}
= \cos\gamma{\sqrt{2}\over
\sqrt{\rho\rho'+\sqrt{\rho^2-1}\sqrt{\rho'{}^2-1}+1}}\,.
\end{split}\ee
It is convenient to use the identity
\be
\mathbf{F}\left(\phi,k\right)={1\over k'}\mathbf{F}\left(\psi,i{k\over
k'}\right)
\ee
with
\be
k'=\sqrt{1-k^2}\hh \sin\psi=k'{\sin\phi\over\sqrt{1-k^2\sin^2\phi}} \,,
\ee
or, equivalently,
\be
k'{}^2={\rho\rho'+\sqrt{\rho^2-1}\sqrt{\rho'{}^2-1}
+1-2\cos^2\gamma\over
\rho\rho'-\sqrt{\rho^2-1}\sqrt{\rho'{}^2-1}
+1-2\cos^2\gamma}\,,
\ee
\be\begin{split}
\sin\psi&={1\over \sqrt{1+\alpha^2}}
=\cos\gamma{\sqrt{2}\over
\sqrt{\rho\rho'-\sqrt{\rho^2-1}\sqrt{\rho'{}^2-1}+1}}\,.
\end{split}\ee
One can rewrite the static Green function  as
\be\begin{split}\nonumber
G(x,x')&={1\over 4\pi^2 a^2}
\left({\partial\over \partial \cos\gamma}\right)
\left\{{\sqrt{2}\over\sqrt{\rho\rho'+\sqrt{\rho^2-1}\sqrt{\rho'{}^2-1}
+1-2\cos^2\gamma} } \left [ \mathbf{F}\left(
\psi,\varkappa
\right)+\mathbf{K}\left(\varkappa\right)\right]\right\}\,.
\end{split}\ee
Here
\be
\varkappa=i{k\over k'}=\sqrt{2\sqrt{\rho^2-1}\sqrt{\rho'{}^2-1}
\over\rho\rho'+\sqrt{\rho^2-1}\sqrt{\rho'{}^2-1}+1-2\cos^2\gamma}\,,
\ee
Eventually we obtain the static Green function
\be\begin{split}
G(x,x')&={1\over 4\pi^2 a^2}{1\over(\rho^2-1)^{1/4}(\rho'{}^2-1)^{1/4}}
\left({\partial\over \partial \cos\gamma}\right)
\left\{\varkappa\left[ \mathbf{F}\left(
\psi,\varkappa
\right)+\mathbf{K}\left(\varkappa\right)\right]\right\}\,.
\end{split}\ee
In this representation $0\le\varkappa\le 1$.

%\bibliography{references}{}

\begin{thebibliography}{10}

\bibitem{Unruh:1976fc}
W.~Unruh, {\it {Selfforce on Charged Particles}},  {\em Proc.Roy.Soc.Lond.}
  {\bf A348} (1976) 447--465.

\bibitem{Smith:1980tv}
A.~Smith and C.~Will, {\it {Force on a static charge outside a Schwarzschild
  black hole}},  {\em Phys.Rev.} {\bf D22} (1980) 1276--1284.

\bibitem{Zelnikov:1983}
A.~Zelnikov and V.~P. Frolov, {\it {The influence of gravitation, acceleration,
  and temperature on the self-energy of charged particles}},  {\em Proceedings
  of the Lebedev Physics Institute (in Russian)} {\bf 152} (1983) 96--116.

\bibitem{Zelnikov:1982in}
A.~Zelnikov and V.~Frolov, {\it {Influence of gravitation on the self-energy of
  charged particles}},  {\em Sov. Phys. JETP} {\bf 55} (1982), no.~2 191--198.

\bibitem{Copson01031928}
E.~T. Copson, {\it On electrostatics in a gravitational field},  {\em
  Proceedings of the Royal Society of London. Series A} {\bf 118} (1928),
  no.~779 184--194,
  
[\href{
http://xxx.lanl.gov/abs/http://rspa.royalsocietypublishing.org/content/118/779/184.fu
ll.pdf+html}{{\tt
  http://rspa.royalsocietypublishing.org/content/118/779/184.full.pdf+html}}].

\bibitem{Beach:2014aba}
M.~J.~S. Beach, E.~Poisson, and B.~G. Nickel, {\it Self-force on a charge
  outside a five-dimensional black hole},  {\em Phys. Rev. D} {\bf 89} (Jun,
  2014) 124014.

\bibitem{Frolov:2014gla}
V.~Frolov and A.~Zelnikov, {\it {Charged particles in higher dimensional
  homogeneous gravitational field: Self-energy and self-force}},
  \href{http://xxx.lanl.gov/abs/1407.3323}{{\tt arXiv:1407.3323}}.

\bibitem{Linet:1976sq}
B.~Linet, {\it {Electrostatics and magnetostatics in the Schwarzschild
  metric}},  {\em J.Phys.} {\bf A9} (1976) 1081--1087.

\bibitem{Linet:1977vv}
B.~Linet, {\it {Scalar or electric charge at rest in a black hole space-time}},
   {\em Compt.Rend.Math.} {\bf 284} (1977) 215--217.

\bibitem{Ottewill:2012aj}
A.~C. Ottewill and P.~Taylor, {\it {Static Kerr Green's Function in Closed Form
  and an Analytic Derivation of the Self-Force for a Static Scalar Charge in
  Kerr Space-Time}},  {\em Phys.Rev.} {\bf D86} (2012) 024036,
  [\href{http://xxx.lanl.gov/abs/1205.5587}{{\tt arXiv:1205.5587}}].

\bibitem{Frolov:2012jj}
V.~P. Frolov and A.~Zelnikov, {\it {Scalar and electromagnetic fields of static
  sources in higher dimensional Majumdar-Papapetrou spacetimes}},  {\em
  Phys.Rev.} {\bf D85} (2012) 064032,
  [\href{http://xxx.lanl.gov/abs/1202.0250}{{\tt arXiv:1202.0250}}].

\bibitem{GarciaParrado:2003hu}
A.~Garcia-Parrado and J.~M. Senovilla, {\it {Bi conformal vector fields and
  their applications}},  {\em Class.Quant.Grav.} {\bf 21} (2004) 2153--2178,
  [\href{http://xxx.lanl.gov/abs/math-ph/0311014}{{\tt math-ph/0311014}}].

\bibitem{GómezLobo20061069}
A.~G.-P. Gómez-Lobo, {\it Bi-conformal vector fields and the local geometric
  characterization of conformally separable pseudo-riemannian manifolds i},
  {\em Journal of Geometry and Physics} {\bf 56} (2006), no.~7 1069 -- 1095.

\bibitem{Frolov:2012zd}
V.~P. Frolov and A.~Zelnikov, {\it {Self-energy of a scalar charge near
  higher-dimensional black holes}},  {\em Phys.Rev.} {\bf D85} (2012) 124042,
  [\href{http://xxx.lanl.gov/abs/1204.3122}{{\tt arXiv:1204.3122}}].

\bibitem{Frolov:2012xf}
V.~P. Frolov and A.~Zelnikov, {\it {Classical self-energy and anomaly}},  {\em
  Phys.Rev.} {\bf D86} (2012) 044022,
  [\href{http://xxx.lanl.gov/abs/1205.4269}{{\tt arXiv:1205.4269}}].

\bibitem{Frolov:2012ip}
V.~P. Frolov and A.~Zelnikov, {\it {Anomaly and the self-energy of electric
  charges}},  {\em Phys.Rev.} {\bf D86} (2012) 104021,
  [\href{http://xxx.lanl.gov/abs/1208.5763}{{\tt arXiv:1208.5763}}].

\bibitem{Frolov:2013qia}
V.~P. Frolov, A.~A. Shoom, and A.~Zelnikov, {\it {Self-energy anomaly of an
  electric pointlike dipole in three-dimensional static spacetimes}},  {\em
  Phys.Rev.} {\bf D88} (2013), no.~2 024032,
  [\href{http://xxx.lanl.gov/abs/1303.1816}{{\tt arXiv:1303.1816}}].

\bibitem{DemiansiNovikov1982}
M.~Demianski and I.~Novikov, {\it Electric charge in the kruskal space-time and
  the jeans conjecture},  {\em General Relativity and Gravitation} {\bf 14}
  (1982), no.~12 1115--1130.

\bibitem{FrolovNovikov1998}
V.~Frolov and I.~Novikov, {\em Black Hole Physics: Basic Concepts and New
  Developments (Fundamental Theories of Physics)}, vol.~96 of {\em Fundamental
  Theories of Physics}.
\newblock Kluwer Academic Publishers, Dordrecht, 1~ed., Nov., 1998.

\bibitem{Poisson:2011nh}
E.~Poisson, A.~Pound, and I.~Vega, {\it {The Motion of point particles in
  curved spacetime}},  {\em Living Rev.Rel.} {\bf 14} (2011) 7,
  [\href{http://xxx.lanl.gov/abs/1102.0529}{{\tt arXiv:1102.0529}}].

\bibitem{Camporesi:1990wm}
R.~Camporesi, {\it {Harmonic analysis and propagators on homogeneous spaces}},
  {\em Phys.Rept.} {\bf 196} (1990) 1--134.

\bibitem{Kofman:1983nj}
L.~Kofman and V.~Sahni, {\it {A new selfconsistent solution of the einstein
  equations with one loop quantum gravitational corrections}},  {\em
  Phys.Lett.} {\bf B127} (1983) 197--200.

\bibitem{Ottewill:2012mq}
A.~C. Ottewill and P.~Taylor, {\it {Quantum field theory on the
  Bertotti-Robinson space-time}},  {\em Phys.Rev.} {\bf D86} (2012) 104067,
  [\href{http://xxx.lanl.gov/abs/1209.6080}{{\tt arXiv:1209.6080}}].

\bibitem{FrolovZelnikov:1980}
V.~P. Frolov and A.~I. Zel'nikov, {\it The massless scalar field around a
  static black hole},  {\em Journal of Physics A: Mathematical and General}
  {\bf 13} (1980), no.~9 L345.

\end{thebibliography}
%\bibliographystyle{JHEP}

%\begin{thebibliography}{99}

\providecommand{\href}[2]{#2}\begingroup\raggedright\endgroup

%\end{thebibliography}

\end{document}